\documentclass[a4paper]{article}
\usepackage{Odyssey2024}
\usepackage{amsmath,graphicx,caption,subcaption,placeins,url,todonotes,makecell,stmaryrd}
\usepackage{float}
\ninept

\title{Automatic Voice Identification after Speech Resynthesis using PPG}

\makeatletter
\def\name#1{\gdef\@name{#1\\}}
\makeatother
\name{{\em Thibault Gaudier$^{1,2}$, Marie Tahon$^1$, Anthony Larcher$^1$, Yannick Estève$^2$}}

\address{$^1$ LIUM / Le Mans Université  \\
$^2$ LIA / Avignon Université\\
{\small \tt \{first\}.\{last\}@univ-lemans.fr} }
\begin{document}
\maketitle

\begin{abstract}

Speech resynthesis is a generic task for which we want to synthesize audio with another audio as input, which finds applications for media monitors and journalists.
Among different tasks addressed by speech resynthesis, voice conversion preserves the linguistic information while modifying the identity of the speaker, and speech edition preserves the identity of the speaker but some words are modified.
In both cases, we need to disentangle speaker and phonetic contents in intermediate representations.
Phonetic PosteriorGrams (PPG) are a frame-level probabilistic representation of phonemes, and are usually considered speaker-independent.
This paper presents a PPG-based speech resynthesis system.
A perceptive evaluation assesses that it produces correct audio quality.
Then, we demonstrate that an automatic speaker verification model is not able to recover the source speaker after re-synthesis with PPG, even when the model is trained on synthetic data.

\end{abstract}

\section{Introduction} \label{sec:intro}

Nowadays media monitors and journalists have to deal with huge content streams from all over the world in different languages. 
In this context, SELMA project\footnote{\url{https://selma-project.eu/}} aims to develope a voice-over framework which will generate a speech signal targeting the voice of a specific journalist/presenter from input translated text.
One option is to use a Text-to-Speech (TTS) system to generate the speech signal corresponding to the translated text expressed by the target voice.
However, despite many recent developements, the signal synthesized with such an approach does not necessarily correspond to what an editor wants.
Therefore, there is a need for the creation of a speech resynthesis framework which enables an expert to directly modify an existing audio file.
To do so, the expert needs to control different aspects of the speech signal generation through the use of an interpretable representation. 
For instance, EditTTS~\cite{edittts} uses text as a representation that permits control over the linguistic content, and \cite{zhao2019ForeignAC} uses Phonetic PosteriorGrams (PPG) as a finer representation to edit rhythmic or phonetic contents.
PPG, as a time-vs-phoneme representation representing posterior probabilities of phonetic classes at the frame level, is the representation we investigate in this paper.

PPG has the advantage of disentangling phonetic and rhythmic information, and thus giving control to our expert on these two aspects independently \cite{zhao2019ForeignAC,cheng2018rhythm}.
PPG has been used for the task of voice conversion \cite{tacovc} (modifying speaker).
Such representation also embeds speaker accent, thus allowing to perform accent conversion \cite{zhao2019ForeignAC} (modifying accent but not the speaker).
However, this means that some potentially unwanted source speaker information could leak from the source audio to the synthetic audio.
Our general objective is to generate a speech signal expressed by a target speaker by resynthesizing an audio file from a PPG. 
Therefore we need to verify that the source speaker cannot be retrieved from the resynthesized audio file.

\begin{figure*} [ht!]
    \centering
    \includegraphics[scale=0.5]{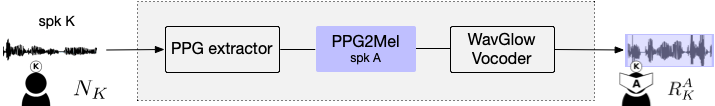}
    \caption{Overview of our PPG-based re-synthesis (PPG2Mel) approach. Blue box denotes speaker-specific model. N stands for natural speech, while R stands for re-synthesized speech (see Sec.~\ref{sec:odn})}
    \label{fig:objective}
\end{figure*}

To do so, we train speech synthesis models on PPG inputs, following \cite{zhao2019ForeignAC} and \cite{tacovc}, as detailed in Figure~\ref{fig:objective}.
More precisely a PPG2Mel network is trained to generate a target voice $A$ from PPG. Consequently, at inference time, when a speech signal $N_K$ from a source speaker $K$ passes through our pipeline, the re-synthesized signal $R_K^A$ is converted towards the target voice $A$.
We then perform a subjective evaluation of the general quality of our speech, comparing it to a TTS baseline, a vocoded only baseline and natural speech. 
The objective of the evaluation described in Section~\ref{sec:ppgss}, is to ensure that the synthetic speech generated from PPG is of correct quality.
As our aim is not to design a new speech synthesis system but to use an existing one for another task, we do not seek for any score-based comparison with state-of-the-art approaches.
However we need to ensure a correct audio quality on the synthesized samples.

As far as we know, no study has reported yet how much source speaker identity is captured by PPGs.
Our main contribution is to investigate in what extent an automatic speaker verification (ASV) system is able to retrieve the source speaker in the resynthesize audio file.
This is done by resynthesizing audio files from different source speakers (VoxCeleb \cite{voxceleb,voxceleb2} dataset) using two models trained on two different target voices. Section~\ref{sec:odn} details the contents of the datasets, their use and the notations that we will use.
With this synthetic dataset annotated with source speakers, we are able to train ASV systems with different setups. Section~\ref{sec:svmodels} details our protocol and Section~\ref{sec:experiments} shows the results we got.
Contrary to voice conversion evaluation protocol, we do not intend to check how similar the synthetic speakers are to the two target voices, but we check how much a ASV system can identify source speakers in the synthetic audio, despite the target voice. 

\section{Related work} \label{sec:related}







All the speech generation tasks are generally divided into two steps : one feature predictor generates a mel-spectrogram from the input, and then a vocoder turns the mel-spectrogram into audio. 
Among the existing neural vocoders, WaveNet~\cite{wavenet} and WaveGlow~\cite{waveglow} have been extensively used with Tacotron2, but the recent HifiGan~\cite{hifigan} which provides faster training and inference, is now the most used one.
Also, some end-to-end systems tend to appear, that embed both steps without relying on an internal signal-related representation of speech, such as Vits~\cite{vits}

Speech generation covers a great variety of tasks (among text-to-speech (TTS), voice conversion (VC) and speech edition).
Each of them requires to retain or exclude different aspects of speech, such as linguistic content, speaker identity, prosody, etc., that need to be disentangled in specific representations of speech.
For example, text modality discards pitch and pronounciation while PPG also seem to discard pitch but are embedding the pronounciation of the sentence.

\subsection{Speech generation from text}

The first task that comes in mind is Text-to-Speech (TTS), where a written sentence must be generated with a target voice.
Some systems, such as Tacotron2~\cite{Tacotron2} or FastPitch~\cite{fastpitch} have been developped for this task in particular.
The easiest case for this task is where only one target voice is used: the voice from the training set. 
Blizzard Challenge~\cite{blizzard2023, blizzard2018} is a TTS challenge, where multiple teams are given one or multiple tasks and a database, and these teams compete to provide the "best" synthetic samples for the tasks. 
The evaluation is done using multiple listening tests in different conditions to order the participants.
Each edition of the challenge uses a different database, which can differ on language (French, Spanish, Mandarin...) or in contents (children book).

Handling multiple voices, or using a speaker representation as a voice target, are harder versions of this task, which is done by other systems. 
This TTS task is usually evaluated with MOS scores, using different questions to capture the opinion of listeners on some precise aspects of speech, such as naturalness or speaker similarity to the target voice.
Some research is also done in the direction of MOS prediction from audio, for example in VoiceMOS Challenge~\cite{voicemos}.

\subsection{Speech generation from audio}

Voice conversion is a speech resynthesis task, in which a target sentence uttered by a source speaker is given, along with a target speaker. The goal is then to generate the target sentence with the target voice, with minimal changes to the aspects which are not the speaker. Evaluation for this task can either be done by running a perceptive test, or by using automatic metrics.
Similarly to the Blizzard Challenge for TTS, the Voice Conversion Challenge is a recurring challenge for Voice Conversion systems.
This challenge has been run every two to three years, starting in 2016.
It provided different tasks over the years to explore different aspects of speech, such as cross-lingual Voice Conversion in 2020~\cite{vcc2020} or Singing Voice Conversion in 2023~\cite{vcc2023}.


Accent conversion is closely related to Voice Conversion. This task consists of changing the accent in the sentence, for example from a non-native to a native accent, without changing the identity nor the words from the original sentence.
In FAC-via-PPG~\cite{zhao2019ForeignAC}, it is done by using PPG as input for a speech synthesis system, here Tacotron2, to modify the pronounciation of the sentence .

Speech anonymization is a variant of Voice Conversion. In this case, the goal is not to generate a specific voice, but instead to not be able to identify the source speaker~\cite{anonymizationSSW}, without interferring with linguistic or prosodic elements.

Speech edition is the last task that we will present. This task consists of, given an audio corresponding to a sentence and a change to do to the audio (the easiest case being replacing a word by another), generating the same audio with the modification taken into account. 
This is done by changing a part of the input text by EditSpeech~\cite{editspeech}, and by changing content or shifting pitch by EdiTTS~\cite{edittts}.

These four tasks, where the criteria can differ from one task to another (speaker similarity, keeping/removing some linguistic elements...) are examples of Speech Resynthesis. 
Compared to TTS, the evaluation could require to compare two audios on some aspects of speech.
This evaluation can be done through perceptive testing or by using automatic metrics in different feature spaces. 

\subsection{Speaker verification and evaluation}

Depending on the targeted task, an evaluation of the similarity of the speakers from two samples can be necessary.
As an example, Voice Conversion aims at maximizing the similarity of the synthetized speaker with a target speaker, but Speech Anonymization suppresses the original speaker identity while maintaining the linguistic content. 
Perceptive evaluation is often done with a Speaker Similarity MOS, where participants to the listening test are asked whether they think that the two presented speakers are the same or not.
Automatic evaluation usually relies on Speaker Verification systems, such as ResNet~\cite{resnet} and Ecapa-TDNN~\cite{ecapatdnn}.
These systems are trained to produce similar embeddings for audios coming from the same speaker, and different embeddings if the speakers are different, regardless of the linguisic contents of the audios.
The input features can be acoustic features, such as MFCC or Mel-Spectrograms, or features extracted by a pretrained model, such as WavLM~\cite{wavlm}.

\subsection{Datasets}
Datasets used for speech generation are usually audiobooks. 
The most common for mono-speaker high-resource English synthesis is LJSpeech~\cite{ljspeech17}. 
This dataset contains 13100 audio segments of 1 to 10 seconds, for a total of 24 hours.
When we want to learn a finite small number of voices, MAILABS~\cite{mailabs} provides, in its English subset, a vast amount of audio for 4 speakers, with 40 to 70 hours per speaker.
This gives the ability to train one-hot speaker-encoded synthesis systems.
This dataset also contains high quantity of audio in 8 other languages, that would give the ability to train speech synthesis systems in these languages.

LibriSpeech~\cite{librispeech} is another dataset, frequently used for many different tasks, such as speech recognition or speaker verification. This dataset contains 960 hours of audio in total, of 2~500 different speakers. 

\begin{figure*} [ht!]
    \centering
    \includegraphics[scale=0.5]{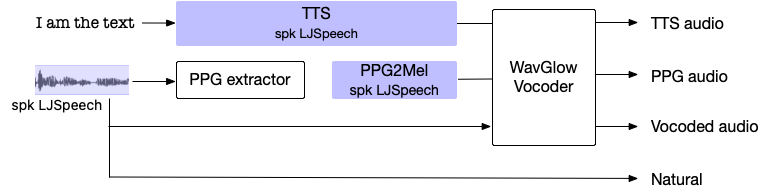}
    \caption{Block representation of the 4 variations of each sample of our perceptive evaluation. 
    Blue boxes are speaker-specific models}
    \label{fig:perceptive}
\end{figure*}

LibriTTS~\cite{libritts} is a subset of LibriSpeech used for Text-to-Speech, and contains 585 hours of read audio.
This gives the possibility to learn a speech synthesis task conditionned on a speaker representation instead of on some learnt voices.

VoxCeleb1\&2 are made of more than 1 million segments of around 7~000 different speakers. They are a common benchmark for the speaker verification task.

\section{Phonetic PosteriorGrams and speech synthesis} \label{sec:ppgss}

\subsection{Phonetic PosteriorGrams (PPG)} \label{sec:ppg}

\begin{figure}[ht]
\centering
\captionsetup{justification=centering}
\includegraphics[width=0.4\textwidth]{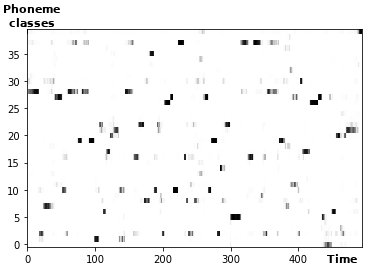}
\caption{PPG example for: "Such risks can be lessened when the President recognizes the security problem"}
\label{fig:ppgexample}
\end{figure}

Phonetic PosteriorGrams (see example Figure~\ref{fig:ppgexample}) are a frame-level representation of speech, which gives a probability of presence of phonemes at each timeframe. This representation can be interesting to give fine-grained control over represented speech to users (see Figure~\ref{fig:objective}), since it disentangles different high-level features, such as pronunciation or rhythm. It also conveys more information than one-hot encoding of phones, since the confusion between two classes can be interpreted as different ways to realize a same phoneme. But some other information might be present in this representation, and this paper aims to look for speaker identity information in a PPG.

Our PPG are extracted from the same model as in \cite{zhao2019ForeignAC}, which is a Kaldi generalized maxout network \cite{maxout} trained to mimic a GMM-HMM model representing 5,816 sub-phone units, which are then grouped into 40 phone classes for English speech. 100 PPG frames are extracted per second.

\subsection{Speech synthesis from PPG (PPG2Mel)} \label{sec:ppg2mel}

Speech generation systems traditionnaly take in input a sequence of characters or phonemes.
PPG are a representation of audio, pretty similar to one-hot encoding of phonemes. Thus, traditional systems can be easily adapted with PPG as an input.
We follow a similar approach as described in~\cite{zhao2019ForeignAC} and~\cite{tacovc}, which involves training a Tacotron2 system~\cite{Tacotron2} using PPG as input, rather than text.

The training should be easier compared to text input because PPG provide a frame-level representation and convey precise timing information. The only modification made to the Tacotron2 architecture are in the first layer of the encoder, where the Character Embedding layer is replaced with a linear layer that maps the 40 phone classes to a 512-sized hidden representation.

\subsection{Mel-Spectrogram to Speech} \label{sec:vocoder}

Since the PPG2Mel system generates a Mel-spectrogram, we need to convert it back to the time domain. To achieve this, we employ WaveGlow, a neural vocoder described in \cite{waveglow}. Our vocoder was trained on the LJSpeech dataset with the default configuration, except for the sampling rate, which we set to 16kHz -- instead of 22.05kHz -- to match the sampling rate of other audio versions. We used the implementation provided by Nvidia, available on their GitHub repository\footnote{\url{https://github.com/nvidia/waveglow}}.

\subsection{Perceptive evaluation of synthetic speech} \label{sec:perceptive}

Speech reynthesis would serve no purpose if it results in a degradation of the synthesized speech quality. Therefore, our initial objective is to assess the quality of the speech generated by the system detailed in Section~\ref{sec:ppg2mel}. In order to ensure that our speech quality remains on par with other speech synthesis methods, we compare our PPG2Mel model with samples generated by a Tacotron2 system that was trained on textual input.

This test does not aim to compare our speech resynthesis system to other on any aspect (naturalness, voice similarity...), since the system we use is not the main contribution of this article, but is an existing system from the literature. The goal of this test is only to make sure the provided audio is of sufficient quality to study the eventual presence of a speaker.


Different audio versions are shown to the listeners, and are summarized in Figure~\ref{fig:perceptive}:
\begin{itemize}
    \item Natural audio : Original audio from LJSpeech dataset, resampled to 16kHz to match the other audios. This will give us the opinion of listeners about natural audio, which is the upper bound for our systems.
    \item Vocoder audio : We extracted mel-spectrograms from original audios, and fed it to the vocoder we use. This gives us the degradation induced by the vocoder, that our synthesis systems will not be able to avoid.
    \item TTS audio : We used a TTS system to synthesize audio from the text given by the dataset. This will be a comparison point for synthetic speech using text or an audio representation such as PPG.
    \item PPG2Mel audio : This is the system we want to test. 
\end{itemize}

Both the PPG2Mel and TTS systems were exclusively trained on the monospeaker LJspeech dataset. As a TTS baseline, we employed Nvidia's implementation of Tacotron2\footnote{\url{https://github.com/nvidia/tacotron2}}, adjusting the sampling rate to 16kHz for consistency with other setups.

For this experiment, we employed the train/dev/test splits from the same repository. We then further divide the test set into three equally sized parts, according to audio duration. We finally selected 20 random segments from each of the three parts to have a representation of short, average and long sentences. We conducted a Mean Opinion Score (MOS) evaluation of these samples to assess their quality. Participants were presented with samples from all four configurations in a randomized manner.

For our perceptive evaluation, we use the FlexEval platform~\cite{flexeval}, which includes a 5-level full-point MOS evaluation page (Bad - Poor - Fair - Good - Excellent). 
Users are asked to "judge the quality of the following sample" and the Welcome page states that "If the overall quality of the samples are very close, you can take into account the expressivity of the samples for your evaluation".

Our test was conducted during 3 weeks, and has mostly been shared across some non-native english speakers from the speech scientific community, about half of them are psycholinguistics students. 
Participants had to answer to 20 steps made of the 4 different variations of one randomly-selected sample. 
The first step is marked as an introduction step for people to familiarize with the test. 44 participants answered to at least two steps, 36 of them completed the whole test and participants answered to 16 steps on average. Each of the 60 samples has been seen approximately 12 times.

\subsection{Results of perceptive evaluation} \label{sec:perceptiveresults}

Results are reported in Table~\ref{tab:mos}. 
We exclude the introduction steps and take into account all the other answers, including those coming from participants who did not complete all the steps.
From these results, we can conclude that using PPG as an input for speech synthesis does not degrade audio quality compared to TTS. 
We also see that a large part of the degradation in audio quality comes from the vocoder, which means that a better training setup or the use of another vocoder could benefit to speech quality of both TTS and PPG2Mel systems.

We are aware tht our results are below similar MOS reported in the litterature. However, we notice that even the natural audio is not evaluated with good score. This states that our participants were particularly strict during the evaluation process compared to state of the art MOS evaluations.

\begin{table}[ht]
\centering
\caption{MOS Scores obtained on our experiment. Confidence intervals at 95\%}
\label{tab:mos}
\begin{tabular}{|c|c|}
\hline
System        & MOS Score \\
\hline
Natural audio & $4.35 \pm 0.07$  \\
Vocoded audio & $3.47 \pm 0.07$ \\
TTS audio     & $3.11 \pm 0.07$ \\
PPG audio     & $3.24 \pm 0.07$ \\
\hline
\end{tabular}

\end{table}

\section{Speaker verification experiment} \label{sec:expdata}

Now that we confirmed the correct audio quality of the resynthesis, we want to perform source speaker verification after PPG-based speech re-synthesis. Since our goal is not voice conversion, the synthetic samples are not required to sound like the target speaker. \\ 

In this section, our objective is to determine whether a Naive Automatic Speaker Verification (ASV) system, trained on natural speech, can successfully identify the source speaker after resynthesis (Q1). 
Subsequently, we employ synthetic data to train an Informed ASV system, designed to recognize the source speaker after resynthesis. We investigate the extent to which this Informed ASV system can recognize the \textbf{source speaker} in samples synthesized with a target voice but also from natural speech samples in order to evaluate the mismatch between natural and synthetic speech (Q2).
The Informed ASV is supposed to learn how to discriminate speakers in a feature space adapted to the target voice.
In case the synthesis process completely hides the source speaker, we expect strong degradations with both Naive and Informed models. 
However, in case the synthesis process only partially hides the source speaker, we expect strong degradatation for the Naive system, and lower degradation for the Informed one. 
Finally, we investigate how much both Naive and Informed systems are able to link the \textbf{target speaker} identity from natural samples and samples synthesized with its target voice (Q3).

\subsection{Data and notations} \label{sec:odn}

\begin{table*}[ht]
    \centering
    \caption{Description of the notations of the different speakers from the datasets we used. Original speakers are noted in a circle above the head. The mask in front of a speaker means that the sample has been synthetized using the voice indicated on the mask. $K$ and $K'$ are different.}
    
    \begin{tabular}{|c|c|c|c|}
        \hline
        \textbf{Dataset} & \textbf{Number of speakers} & \textbf{Notation} & \textbf{Details}\\
        \hline
        M-AILABS & 4 & 
        \begin{minipage}{.065\textwidth}
            \includegraphics[scale=0.8]{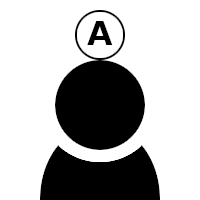}
        \end{minipage}
        \begin{minipage}{.065\textwidth}
            \includegraphics[scale=0.8]{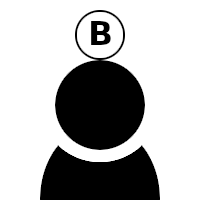}
        \end{minipage}
        & $A$ is E. Klett, $B$ is E. Miller\\
        \hline
        LibriSpeech & 40 & 
        \begin{minipage}{.065\textwidth}
            \includegraphics[scale=0.8]{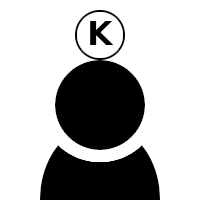}
        \end{minipage}
        \begin{minipage}{.065\textwidth}
            \includegraphics[scale=0.8]{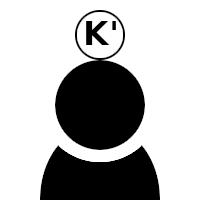}
        \end{minipage}
        & $K, K' \in \llbracket 1, 40 \rrbracket, K \neq K'$ \\
        \hline
        Synthetic LibriSpeech-test & 40 & 
        \begin{minipage}{.065\textwidth}
            \includegraphics[scale=0.8]{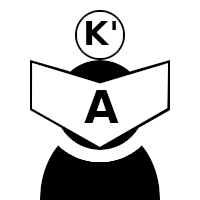}
        \end{minipage}
        \begin{minipage}{.065\textwidth}
            \includegraphics[scale=0.8]{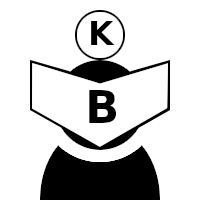}
        \end{minipage}
        & \makecell{$K, K' \in \llbracket 1, 40 \rrbracket, K \neq K'$\\ $A, B$ as described in M-AILABS} \\
        \hline        
        VoxCeleb1\&2 & 7363 & 
         \begin{minipage}{.065\textwidth}
            \includegraphics[scale=0.8]{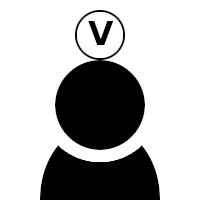}
        \end{minipage}
        \begin{minipage}{.065\textwidth}
            \includegraphics[scale=0.8]{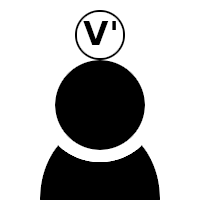}
        \end{minipage}
        & Only used for training naive model\\
        \hline
        Synthetic VoxCeleb1\&2 & 7363 & 
        \begin{minipage}{.065\textwidth}
            \includegraphics[scale=0.8]{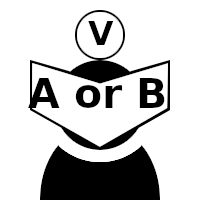}
        \end{minipage}
        \begin{minipage}{.065\textwidth}
            \includegraphics[scale=0.8]{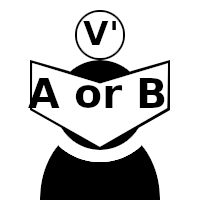}
        \end{minipage} 
        & \makecell{Only used for training Informed ASV\\ Synthesized using speakers $A$ and $B$ from M-AILABS.}\\
        \hline

    \end{tabular}
    \label{tab:dataset}
\end{table*}

This experiment is realized upon 3 different datasets. The first one is the English section of M-AILABS \footnote{\url{https://www.caito.de/2019/01/03/the-m-ailabs-speech-dataset/}}, a read speech corpus based on LibriVox. 
We used two speakers, E. Klett, denoted Speaker $A$ (female), and E. Miller, denoted Speaker $B$ (male), as described in Table~\ref{tab:dataset}, row 1.
Each of these speakers provided 30 to 45 hours of speech data, which we divided into training, validation, and test sets.
Throughout this experiment, those two speakers consistently served as the \textbf{target speakers}. This means that all synthetic samples used in this experiment were converted to one of these two voices.
We trained two mono-speaker PPG2Mel$_A$ and PPG2Mel$_B$ models, one for each speaker $A$ and $B$ (Fig.~\ref{fig:objective}). 

LibriSpeech-test-clean \cite{librispeech} subset is our speaker verification enrollment and test sets.
It contains 40 gender-balanced speakers ($\simeq$ 8 min speech), denoted Speaker $1$ to $40$, as shown in Table~\ref{tab:dataset}, row 2.
These speakers are the \textbf{source speakers} we want to recognize with ASV systems before/after resynthesis.

We create two synthetic versions of this subset, using PPG2Mel$_A$ and PPG2Mel$_B$, shown in row 3 of Table~\ref{tab:dataset}.

Finally, VoxCeleb1\&2 \cite{voxceleb, voxceleb2} datasets are used to train ASV models (row 4 of Table~\ref{tab:dataset}). 

The two mono-speaker PPG2Mel$_A$ and PPG2Mel$_B$ models are used to convert all samples from VoxCeleb towards a target speaker, randomly chosen among $A$ and $B$ (row 5 of Table~\ref{tab:dataset}.
The speaker labels for training remain the same as in the original dataset, even if they now have a different voice.

Natural samples are denoted $N_{source}$, where $source$ is within $\{A, B,1-40\}$. Synthetic samples are denoted $R_{source}^{target}$, where $source$ is as previously mentionned and $target$ is either $A$ or $B$ depending on the PPG2Mel model used. We will not mention the speakers of VoxCeleb datasets.

\subsection{ASV models} \label{sec:svmodels}

Both Naive and Informed ASV models use an ECAPA-TDNN \cite{ecapatdnn} architecture fed with input features obtained by processing the speech samples with a WavLM-large pretrained model\footnote{\url{https://github.com/microsoft/unilm/tree/master/wavlm}} and trained using an AAM loss. 256-dimension x-vectors are extracted.

The Naive ASV model is trained on the original (natural) VoxCeleb1\&2 development data.

This model achieved 1.57\% EER on VoxCeleb-o after 4 days of training on one RTX8000 GPU.
The Informed ASV model is trained on the synthetic version (target speaker $A$ or $B$) of VoxCeleb1\&2 data described in the previous section. The best version of this system is obtained after 1 day on the same architecture and only obtains 20\% of EER on the synthetic version of the VoxCeleb-o task.

\subsection{Experiments and results} \label{sec:experiments}

\begin{table*}[ht!]
    \centering
    \renewcommand{\arraystretch}{1.8}
    \caption{Equal Error rates obtained with both Naive and Informed Speaker Verification models. For each experiment, test definitions are given as \textit{enrollement/test} where $N$ and $R$ refers to natural and resynthesised speech samples respectively. A test defined as  $N_{1-40}/N_{1-40}$ compares couples of natural samples from speakers $\{1-40\}$, for instance $N_1 / N_1$ while a test defined as 
    $N_{1-40}/R_{\overline{1-40}}^{A}$ compares natural speech samples from all speakers ${1-40}$ with resynthesised samples from other speakers 
    among $\{1-40\}$, for instance $N_1 / R^A_2$.}
    \begin{tabular}{|c|c|c|c|c|c|}
        \hline
        Experiment ID    & (1)          & (2)            & (3)              & (4)           & (5)\\
        \hline
        Target test definition        & $N_{1-40}/N_{1-40}$   & $R_{1-40}^{A}/R_{1-40}^{A}$ & $N_{1-40}/R_{1-40}^{A}$       & $N_{A}/R_{1-40}^A$       & $N_{A}/R_{1-40}^A$ \\
        \hline
        Target test definition
        &
        \begin{minipage}{.065\textwidth}
            \includegraphics[scale=0.8]{figures/persos/human_k.png}
        \end{minipage}
        \begin{minipage}{.065\textwidth}
            \includegraphics[scale=0.8]{figures/persos/human_k.png}
        \end{minipage}
        &
        \begin{minipage}{.065\textwidth}
            \includegraphics[scale=0.8]{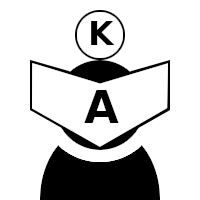}
        \end{minipage}
        \begin{minipage}{.065\textwidth}
            \includegraphics[scale=0.8]{figures/persos/human_k_a.png}
        \end{minipage}
        &
        \begin{minipage}{.065\textwidth}
            \includegraphics[scale=0.8]{figures/persos/human_k.png}
        \end{minipage}
        \begin{minipage}{.065\textwidth}
            \includegraphics[scale=0.8]{figures/persos/human_k_a.png}
        \end{minipage}
        &
        \begin{minipage}{.065\textwidth}
            \includegraphics[scale=0.8]{figures/persos/human_a.png}
        \end{minipage}
        \begin{minipage}{.065\textwidth}
            \includegraphics[scale=0.8]{figures/persos/human_k_a.png}
        \end{minipage}
        &
        \begin{minipage}{.065\textwidth}
            \includegraphics[scale=0.8]{figures/persos/human_a.png}
        \end{minipage}
        \begin{minipage}{.065\textwidth}
            \includegraphics[scale=0.8]{figures/persos/human_k_a.png}
        \end{minipage}
        \\
        \hline
        Impostor test definition       & $N_{1-40}/N_{\overline{1-40}}$   & $R_{1-40}^{A}/R_{\overline{1-40}}^{A}$ & $N_{1-40}/R_{\overline{1-40}}^{A}$   & $N_A/N_{1-40}$ & $N_A / R_{1-40}^{B}$ \\
        \hline
                Impostor test definition
        &
        \begin{minipage}{.065\textwidth}
            \includegraphics[scale=0.8]{figures/persos/human_k.png}
        \end{minipage}
        \begin{minipage}{.065\textwidth}
            \includegraphics[scale=0.8]{figures/persos/human_k2.png}
        \end{minipage}
        &
        \begin{minipage}{.065\textwidth}
            \includegraphics[scale=0.8]{figures/persos/human_k_a.png}
        \end{minipage}
        \begin{minipage}{.065\textwidth}
            \includegraphics[scale=0.8]{figures/persos/human_k2_a.png}
        \end{minipage}
        &
        \begin{minipage}{.065\textwidth}
            \includegraphics[scale=0.8]{figures/persos/human_k.png}
        \end{minipage}
        \begin{minipage}{.065\textwidth}
            \includegraphics[scale=0.8]{figures/persos/human_k2_a.png}
        \end{minipage}
        &
        \begin{minipage}{.065\textwidth}
            \includegraphics[scale=0.8]{figures/persos/human_a.png}
        \end{minipage}
        \begin{minipage}{.065\textwidth}
            \includegraphics[scale=0.8]{figures/persos/human_k.png}
        \end{minipage}
        &
        \begin{minipage}{.065\textwidth}
            \includegraphics[scale=0.8]{figures/persos/human_a.png}
        \end{minipage}
        \begin{minipage}{.065\textwidth}
            \includegraphics[scale=0.8]{figures/persos/human_k_b.png}
        \end{minipage}
        \\
        \hline
        Naive ASV        & 1.98 \%       & 49.46 \%        & 48.02 \%          &     45.13 \%  &  49.81\% \\
        Informed ASV     & 29.44 \%      & 49.80 \%        & 49.00 \%          &     33.58 \%  &  45.52\% \\
        \hline
    \end{tabular}
    \label{tab:eer}
\end{table*}

Table~\ref{tab:eer} summarizes the different set up and results obtained for the 5 speaker verification experiments. Each experiment is described with corresponding enrollement/test pairs.
Experiments come with Equal Error Rates (EER) calculated with both Naive and Informed ASV models. 
For all tests in experiments (1), (2) and (3), the speaker reference labels are the one from speakers 1-40 while in experiments (4) and (5) the labels are $A$ or $B$.
For example, in exp (3), enrollement is done with natural samples from speakers 1-40, while test is done with synthetic samples from speakers 1-40 converted with PPG2Mel$_A$. Target pairs are 1-1, 2-2,~$\ldots$ 40-40, while impostor pairs are 1-2, 1-3, $\ldots$ 39-40.

The first experiment (1) ensures that our naive model achieves correct results. To do so, we want to recover the speaker from natural speech. 
As we could expect, the naive model gives good results (EER=1.98\%) since it is the task it has been trained on, and the informed model introduces a strong degradation (EER=29.44\%), showing that there is a mismatch between the training data used for this model and the test data.

Experiment (2) tests naive and informed models on synthetic speech. We compare samples from the speakers $1-40$, which were all synthetized using PPG2Mel$_A$ ($R_{1-40}^A$), and we want to see if our models are able to recognize those which come from the same source speaker. 
The results show that both naive and informed ASV mocels are unable to discriminate source speaker in the synthetic speech space (EER $> 49\%)$. 
This answers to the question Q1: the naive model is not able to recognize the source speaker after re-synthesis (EER=49.46\%)
One hypothesis is that source speaker identities have been hidden during re-synthesis.
We can see that the informed model better recognizes speakers in the natural space (EER=29.44\%, exp. (1)) than in the synthetic space (EER=49.80\%, exp. (2)).
During training, the informed model hardly converged, but it seems that the few it learnt enables to discriminate speakers in the natural space only (as the task is easier). 
The answer to the question Q2 is: even an informed model is not able to recognize the source speaker after re-synthesis. 

Experiment (3) assesses the ability of both models to make the link between the speakers in the natural space and the same speakers in the synthetic space. To do so, we use natural LibriSpeech dataset as our enrollment dataset and the synthetic version of the same dataset, using PPG2Mel$_A$, as our test data. 
We see that both naive and informed models are unable to recognize the 40 source speakers after re-synthesis.
We conclude that our PPG approach is indeed able to hide source speaker identity, even to an informed ASV.
Any source speaker acoustic clues which could help model to retrieve their identity is not detected after re-synthesis. 

Experiment (4) and (5) measure if source speakers $1-40$ converted in $A$ synthetic space are closer to their natural version (resp. synthetic version converted in $B$) than to the natural speech of $A$.
From experiment (4), we conclude that the speaker identity of the samples from source speaker $1-40$, re-synthesized with target speaker $A$, does not correspond to the identity of $A$, thus confirming the fact that we are not doing voice conversion.
However, the results show that re-synthesized samples are closer to natural samples of speaker $A$ with the informed model than with the naive model.
We thus confirm that for the informed ASV, re-synthesis seems to bring synthetic and natural identity closer.

Experiment (5) shows that synthetic samples generated with target speakers $A$ or $B$ are not distinguishable by the naive system, and both are far from speaker $A$. The informed system makes a subtle difference between samples generated in the spaces of speaker $A$ and $B$.
Therefore our framework is clearly not doing voice conversion and the answer to question Q3 is that the informed model is slighly better than the naive one in finding the link between the target speaker identity from natural samples and samples synthesized with its target voice.
However, these results must be handle with care as this last experiment has been conducted on two target speakers only.

\section{Conclusion} \label{sec:conclusion}

Our first experiment aims to ensure that the PPG2Mel model we trained produced audio of a correct quality. The perceptual study that we performed confirmed that we were achieving the same quality as our TTS baseline, and that the vocoder was producing most of the quality loss.



We trained a naive ASV system on natural speech and an informed ASV system on synthetic speech to try to recover source speaker information hidden by speech synthesis. Our experiments show that even if the naive model achieves state of the art results on natural speech, neither naive nor informed ASV systems are able to retrieve source speaker information which would come from the PPG. This implies that source speaker acoustic clues are not detected by the models in re-synthetized speech. Also, we showed that both ASV models are not able to link target speakers from natural and synthetic samples.

These speaker identification results show that the amount of speaker information that goes from the PPG to the synthesized sample is small enough to permit the use of PPG in tasks such as Voice Conversion or Speech Edition.

Future work would include using a better vocoder, and running a speaker similarity perceptive evaluation of our systems, to compare our approach to Voice Conversion models and to the results obtained through automatic speaker verification.

Since PPG do not convey acoustic speaker clues, we advocate for their use in speech edition, as a speech controllable representation without biasing the ouput towards source speaker.

\section{Acknowledgements}

This work was performed using HPC resources from GENCI–IDRIS (Grants 2022-AD011012565 and AD011012527). This project has also received funding from the European Union's Horizon 2020 research and innovation program under the Marie Skłodowska-Curie grant agreement No 101007666. This paper was partially funded by the European Commission through the SELMA project under grant number 957017.

\bibliographystyle{IEEEbib}
\bibliography{interspeech/mybib}

\end{document}